\documentstyle[twocolumn,graphicx,aps]{revtex}

\begin{document}
\draft

\title{Superfluidity vs Bose-Einstein condensation in a Bose gas with disorder}

\author{G.E. Astrakharchik$^{1}$, J. Boronat$^{2}$, J. Casulleras$^{2}$ and S. Giorgini$^{1}$}

\address{
{\small\it  $^{1}$Dipartimento di Fisica, Universit\`a di Trento, and Istituto 
Nazionale per la Fisica della Materia, I-38050 Povo, Italy}\\
{\small\it $^{2}$Departament de F\'{\i}sica i Enginyeria Nuclear, Campus Nord B4-B5, 
Universitat Polit\`ecnica de Catalunya, E-08034 Barcelona, Spain}
\\ (\today)
\\ \medskip}\author{\small\parbox{14.2cm}{\small\hspace*{3mm}
We investigate the phenomenon of Bose-Einstein condensation and superfluidity in a Bose gas at zero 
temperature with disorder. By using the Diffusion Monte-Carlo method we calculate the superfluid and
the condensate fraction of the system as a function of density and strength of disorder. 
In the regime of weak disorder we find agreement with the analytical results obtained within the Bogoliubov 
model. For strong disorder the system enters an unusual regime where the superfluid fraction is smaller than 
the condensate fraction.\\
\\[3pt]PACS numbers: 03.75.Fi, 05.30.Fk, 67.40.Db}} \maketitle

\narrowtext

The study of disordered Bose systems has attracted in the recent past considerable attention both
theoretically and experimentally.
The problem of boson localization, the superfluid-insulator transition and the nature of  
elementary excitations in the presence of disorder have been the object of several theoretical
investigations \cite{DISTH} and Monte-Carlo numerical simulations \cite{DISMC1,DISMC2}, both based 
on Hubbard or equivalent models on a lattice. More recently, the problem of Bose systems with 
disorder has also been addressed in the continuum. On the one hand, the dilute Bose gas with disorder has been 
studied within the Bogoliubov model \cite{Huang,GPS,Vinokur}. On the other, Path Integral Monte-Carlo 
(PIMC) techniques have been applied to the study of the elementary excitations in liquid $^4$He 
\cite{Boninsegni} and the transition temperature of a hard-sphere Bose gas \cite{Gordillo}, 
in the presence of randomly distributed static impurities. Disordered Bose systems are produced 
experimentally in liquid $^4$He adsorbed in porous media, such as Vycor or silica gels (aerogel,
xerogel). The suppression of superfluidity and the critical behavior at the phase transition have 
been investigated in these systems in a classic series of experiments \cite{Reppy}, and the 
elementary excitations of liquid $^4$He in Vycor have been recently studied using neutron inelastic
scattering \cite{Glyde}. Furthermore, the recent achievement of Bose-Einstein condensation (BEC) in 
alkali vapours has sparked an even larger interest in the physics of degenerate Bose gases and their 
macroscopic quantum properties, such as long-range order and superfluid behavior 
(for a review see \cite{DGPS}). 

In this Letter we investigate the effects of disorder on BEC and superfluidity in a Bose gas at zero 
temperature. As a model for disorder a uniform random distribution of static impurities is assumed. 
This choice provides us with a reasonable model for $^4$He adsorbed in porous media and might also be 
relevant for trapped Bose condensates in the presence of heavy impurities. In addition, the quenched-impurity 
model allows us to derive analytical results in the weak-disorder regime and can be implemented
in a quantum Monte Carlo simulation. 

The present work is divided in two parts. In the first part, 
following the analysis of Ref. \cite{Huang}, the properties of the system are investigated within the Bogoliubov 
approximation. Results for the effects of disorder on the ground-state energy, superfluid 
density and condensate fraction are discussed. In the second part, we resort to the Diffusion Monte-Carlo (DMC) 
technique, which solves exactly the many-body Schr\"odinger equation for the ground state of a boson system. 
By using this technique, we verify that the results of the Bogoliubov model apply only to 
dilute systems with weak disorder and we investigate the cross-over to the regime of strong disorder, 
where the suppression of superfluidity and BEC due to the random potential is large. In this regime we 
find that the system exhibits the unusual feature of a superfluid component smaller than
the condensate component. 

{\it Bogoliubov model.} The starting point is the Bogoliubov Hamiltonian of a homogeneous dilute 
Bose gas 
\begin{equation}
H_0=E_0 + \sum_{\bf p} \epsilon_p \alpha_{\bf p}^\dagger\alpha_{\bf p} \; ,
\label{bog}
\end{equation}
written in terms of the quasi-particle annihilation and creation operators $\alpha_{\bf p}$, 
$\alpha_{\bf p}^\dagger$. These operators are related to the particle operators $a_{\bf p}$, $a_{\bf p}^\dagger$ 
through the well-known canonical transformation $a_{\bf p}=u_p \alpha_{\bf p} + v_p \alpha_{- {\bf p}}^\dagger$, 
with coefficients $u_p^2=1+v_p^2=(\epsilon_p^0+gn_0+\epsilon_p)/2\epsilon_p$ and $u_pv_p=-gn_0/2\epsilon_p$. 
The elementary excitation energies obey the usual Bogoliubov spectrum 
$\epsilon_p=[(\epsilon_p^0)^2+2gn_0\epsilon_p^0]^{1/2}$, with $\epsilon_p^0=p^2/2m$ the free particle energy, 
$n_0$ the condensate density and $g=4\pi\hbar^2a/m$ the coupling constant fixed by the $s$-wave scattering length $a$. 
The constant term    
$E_0/N=[4\pi na^3 + 512\sqrt{\pi}(na^3)^{3/2}/15]\hbar^2/(2ma^2)$ is the ground-state energy per particle expressed 
in terms of the gas parameter $na^3$, with $n=N/V$ the total particle density.
This result includes the zero-point motion of the elementary excitations.

Disorder is introduced in the system by adding to $H_0$ the perturbation 
$H^\prime=\int d^3{\bf r}\; V({\bf r})n({\bf r})$ produced 
by the external field $V({\bf r})=\sum_{i=1}^{N_{\rm imp}}v(|{\bf r}-{\bf r}_i|)$ associated with the impurities.
Here, $N_{\rm imp}$ counts the impurities with fixed position ${\bf r}_i$ and $v(r)$ is the two-body 
particle-impurity potential. For dilute systems and small concentrations of impurities the pair 
potential $v(r)$ can be expressed
as a pseudo-potential $v({\bf r})=g_{\rm imp}\delta({\bf r})$. The coupling constant $g_{\rm imp}=2\pi\hbar^2b/m$  
is fixed by the particle-impurity $s$-wave scattering length $b$ and by the reduced mass of the pair, which coincides 
with the particle mass $m$ if the impurity is infinitely massive. Assuming a uniform random distribution 
of impurities with density $n_{\rm imp}=N_{\rm imp}/V$ and gaussian correlated disorder, we obtain that the statistical
properties of disorder are described by the average value $\langle V_0\rangle = 1/V \int d^3{\bf r}\; 
\langle V({\bf r})\rangle = g_{\rm imp}n_{\rm imp}$, and by the correlation function $C({\bf s})=1/V \int d^3{\bf r}\; 
\langle V({\bf r})V({\bf r}+{\bf s})\rangle$, whose Fourier transform is given by 
$\langle V_{\bf p}V_{-{\bf p}}\rangle = 1/V \int d^3{\bf s}\;e^{-i{\bf p}{\bf s}/\hbar}C({\bf s})=
g_{\rm imp}^2n_{\rm imp}/V$. The notation $\langle ..\rangle$ stands here for average over disorder configurations.
The model is described by three parameters: i) the gas parameter $na^3$, ii) the concentration of impurities
$\chi=N_{\rm imp}/N$, and iii) the ratio of scattering amplitudes $b/a$. The first parameter is related to the strength 
of interactions, the other two to the strength of disorder. Within the Bogoliubov model all relevant properties 
of the system depend on disorder through the combination $R=\chi\;(b/a)^2$, which gives a 
measure of the strength of disorder.

The perturbation term $H^\prime$ can be written in momentum space as 
$H^\prime=NV_0+\sum_{\bf p}V_{-{\bf p}}\rho_{\bf p}$, where $\rho_{\bf p}$ is the density fluctuation operator. 
Within the Bogoliubov approximation we write
$\rho_{\bf p}\simeq \sqrt{N_0}(a_{\bf p}+a_{-{\bf p}}^\dagger)=\sqrt{N_0}(u_p+v_p)(\alpha_{\bf p}+
\alpha_{-{\bf p}}^\dagger)$, where $N_0$ is the number of atoms in the condensate. The total Hamiltonian
$H=H_0+H^\prime$ is given by a combination of linear and quadratic terms in the quasi-particle operators 
$\alpha_{\bf p}$, $\alpha_{\bf p}^\dagger$ and can be diagonalized by means of the operator shift  
\cite{Huang} $\alpha_{\bf p}=\beta_{\bf p}-\sqrt{N_0}V_{\bf p}(u_p+v_p)/\epsilon_p$. One finds
\begin{equation}
H=E+\sum_{\bf p} \epsilon_p \beta_{\bf p}^\dagger\beta_{\bf p} \;.
\label{bog1}
\end{equation}
To lowest order, the elementary excitation energies are not affected by the random field, whereas the ground-state
energy is given by $E=E_0+N[g_{\rm imp}n_{\rm imp}-g_{\rm imp}^2n_{\rm imp}(1/V)\sum_{\bf p}2m/(p^2+4mgn_0)]$. 
The term proportional 
to $g_{\rm imp}^2$ is ultraviolet divergent, but the difficulty is overcome if one takes into account the second
order correction to the particle-impurity coupling constant $g_{\rm imp}\to g_{\rm imp} + g_{\rm imp}^2(1/V)\sum_{\bf p}
2m/p^2$. The final result for the ground-state energy per particle in units of $\hbar^2/2ma^2$ reads 
\begin{equation}
\frac{E}{N}=\frac{E_{MF}}{N} + (na^3)^{3/2}
\left[ \frac{512\sqrt{\pi}}{15} + 16\pi^{3/2} \; R \right] \;,
\label{gse}
\end{equation}
where $E_{MF}/N=4\pi (na^3)[1+\chi\;(b/a)]$ is the mean-field contribution. 
Notice that the model of $\delta$-correlated disorder of Refs. 
\cite{Huang,GPS,Vinokur} does not allow the calculation of the ground-state energy, since the renormalization
of $g_{\rm imp}$ is a crucial step. 

The depletion of the condensate and the non-superfluid component of the gas can be obtained 
from the Hamiltonian (\ref{bog1}) by calculating, respectively, the momentum distribution and 
the long-wavelength behavior of the static transverse current-current response function \cite{Huang,GPS}. 
For the condensate fraction one finds 
\begin{equation}
\frac{N_0}{N}=1-\frac{8}{3\sqrt{\pi}}(na^3)^{1/2}-\frac{\sqrt{\pi}}{2}(na^3)^{1/2}\;R \;,
\label{CF}
\end{equation}
in which the first term gives the quantum depletion due to interaction and the second term accounts for the effect of
disorder. Differently from $N_0/N$, the superfluid fraction is equal to unity in the absence of disorder and one has
\begin{equation}
\frac{\rho_s}{\rho}=1-\frac{4}{3}\frac{\sqrt{\pi}}{2}(na^3)^{1/2}\;R \;.
\label{SF}
\end{equation} 
As it has been anticipated, both the result for the energy beyond mean field (\ref{gse}) and results (\ref{CF}) 
and (\ref{SF}) depend on disorder through the scaling parameter $R=\chi\;(b/a)^2$. Another interesting consequence
of the above results is that, due to the coefficient $4/3$ in (\ref{SF}), disorder is more efficient in depleting
the superfluid than the condensate fraction \cite{Huang}. 
In addition, it is predicted that for any value of $na^3$ there exists a critical
strength of disorder $R_c=16/\pi\simeq 5.1$ for which $\rho_s/\rho<N_0/N$. The results of the Bogoliubov 
model are expected to be valid for dilute systems and weak disorder. However, it is not clear whether these results
still apply for $R>R_c$ in a range of densities where the difference between $\rho_s/\rho$ and $N_0/N$ 
can be significant. These questions have been addressed using the DMC method.

{\it DMC simulation.} We consider a system of $N$ spinless bosons of mass $m$ and $N_{\rm imp}$ impurities placed 
at random in a box with periodic boundary conditions. The Hamiltonian of the system is given by
$H=- (\hbar^2/2m)\sum_{i=1}^N\nabla_i^2+\sum_{i<j}u(|{\bf r}_i-{\bf r}_j|)+\sum_{i=1}^N\sum_{\ell=1}^{N_{\rm imp}}
v(|{\bf r}_i-{\bf r}_\ell|)$, where $u(r)$ and $v(r)$ are respectively the particle-particle and particle-impurity
two-body potential. For both potentials we use a hard-sphere model: particles have diameter $a$ and impurities have
diameter $2b-a$, where $b$ is the range of $v(r)$. Impurities have fixed position ${\bf r}_\ell$ and overlap between
impurities is avoided. Importance sampling is used through the trial wavefunction $\psi_T({\bf R}) \equiv 
\psi_T({\bf r}_1,..,{\bf r}_N) = \prod_{i<j}f(r_{ij})\prod_{i,\ell}g(r_{i\ell})$. The Jastrow factors, $f(r)$ of a 
pair of particles and $g(r)$ of a particle-impurity pair, are calculated using the same technique as in 
Ref. \cite{US}. Average over disorder is obtained by repeating the simulation for different configurations of 
impurities. A number between 5 and 10 independent configurations has proven to be enough. The direct
output of the DMC algorithm is the ground-state energy, which is exact apart from statistical uncertainty
(for further details on the DMC method see Ref. \cite{BC}).
The superfluid fraction $\rho_s/\rho$ can be calculated by extending to zero temperature the winding-number 
technique employed in PIMC calculations \cite{Ceperley}, as discussed for bosons on a lattice in 
Ref. \cite{DISMC2}. The superfluid fraction is obtained as the ratio of two diffusion constants 
$\rho_s/\rho=D_s/D_0$, where $D_0=\hbar^2/2m$ is the diffusion constant in imaginary time of a free particle
and 
\begin{equation}
D_s=\lim_{\tau\to\infty}\frac{N}{6\tau}\frac{\int d{\bf R}\;f({\bf R},\tau) [R_{\rm CM}(\tau)-R_{\rm CM}(0)]^2}
{\int d{\bf R}\;f({\bf R},\tau)} \;,
\label{supfrac}
\end{equation}
is the diffusion constant of the ``center of mass" of the system $R_{\rm CM}=(1/N)\sum_{i=1}^N {\bf r}_i$. 
In the above 
equation $f({\bf R},\tau)$ is the probability density of walkers generated by the DMC algorithm during integration in  
imaginary time $\tau$. One can prove that the above result for $\rho_s/\rho$ is exact and does not
depend on the choice of the trial wavefunction \cite{Astra}. Finally, the condensate fraction is obtained from the 
long-range
behavior of the one-body density matrix: $N_0/N=\lim_{r\to\infty}\rho(r)$ (see Ref. \cite{BC} for further details).
We performed calculations for values of $N =$ 16, 32 and 64 and no significative finite-size effects were found.

\begin{figure}
\begin{center}
\includegraphics*[width=0.95\columnwidth,height=0.6\columnwidth]{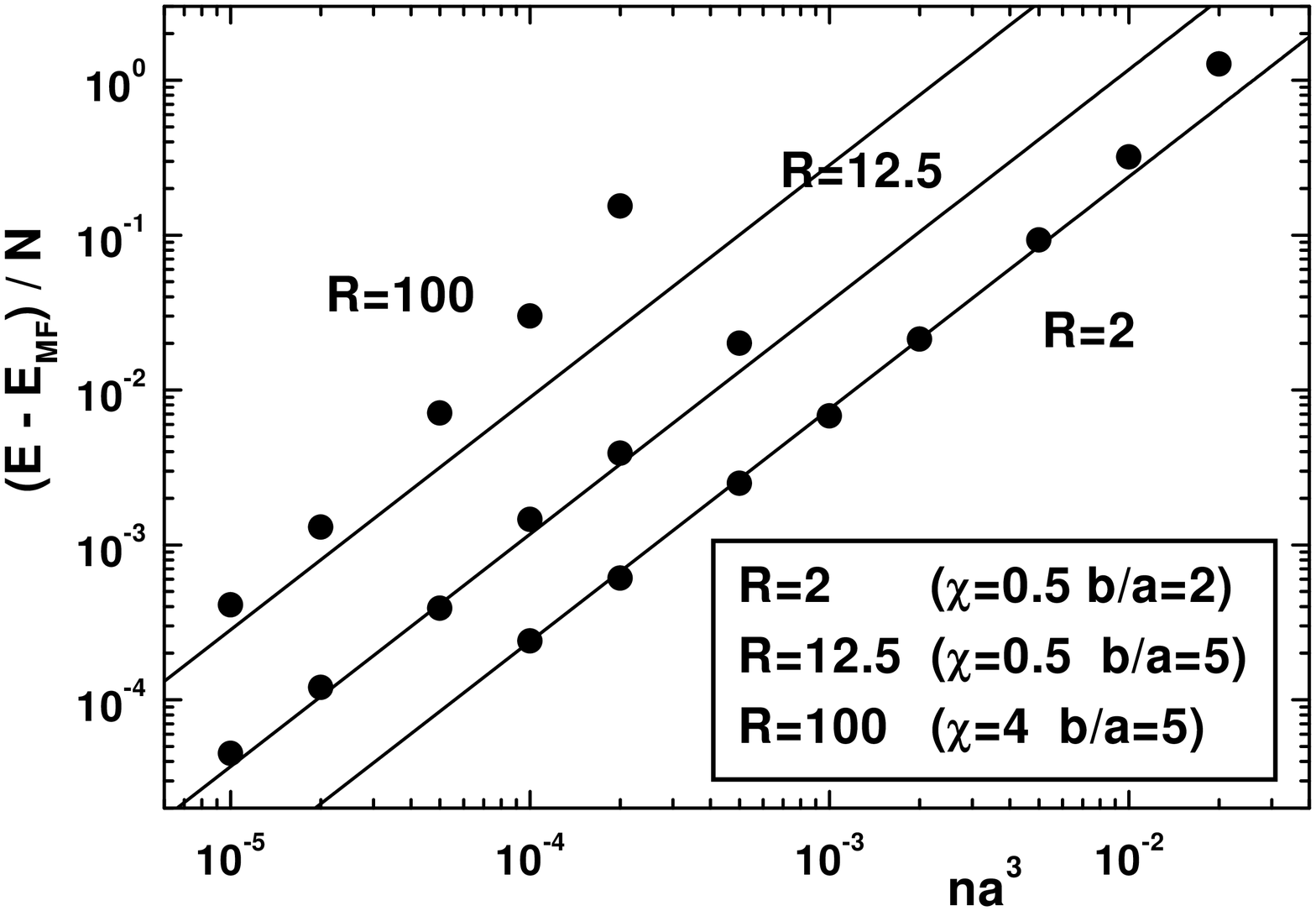}
\vspace{3 mm}
\caption{Energy per particle beyond mean field. The results for a given strength of disorder $R$ are obtained 
for a fixed concentration $\chi$ and a fixed ratio $b/a$ as shown in the figure. The error bars are smaller than
the size of the symbols. The solid lines correspond to Eq. (\ref{gse}). Energies are in units of $\hbar^2/2ma^2$.}
\label{Fig1}
\end{center}
\end{figure}

\begin{figure}
\begin{center}
\includegraphics*[width=0.95\columnwidth,height=0.6\columnwidth]{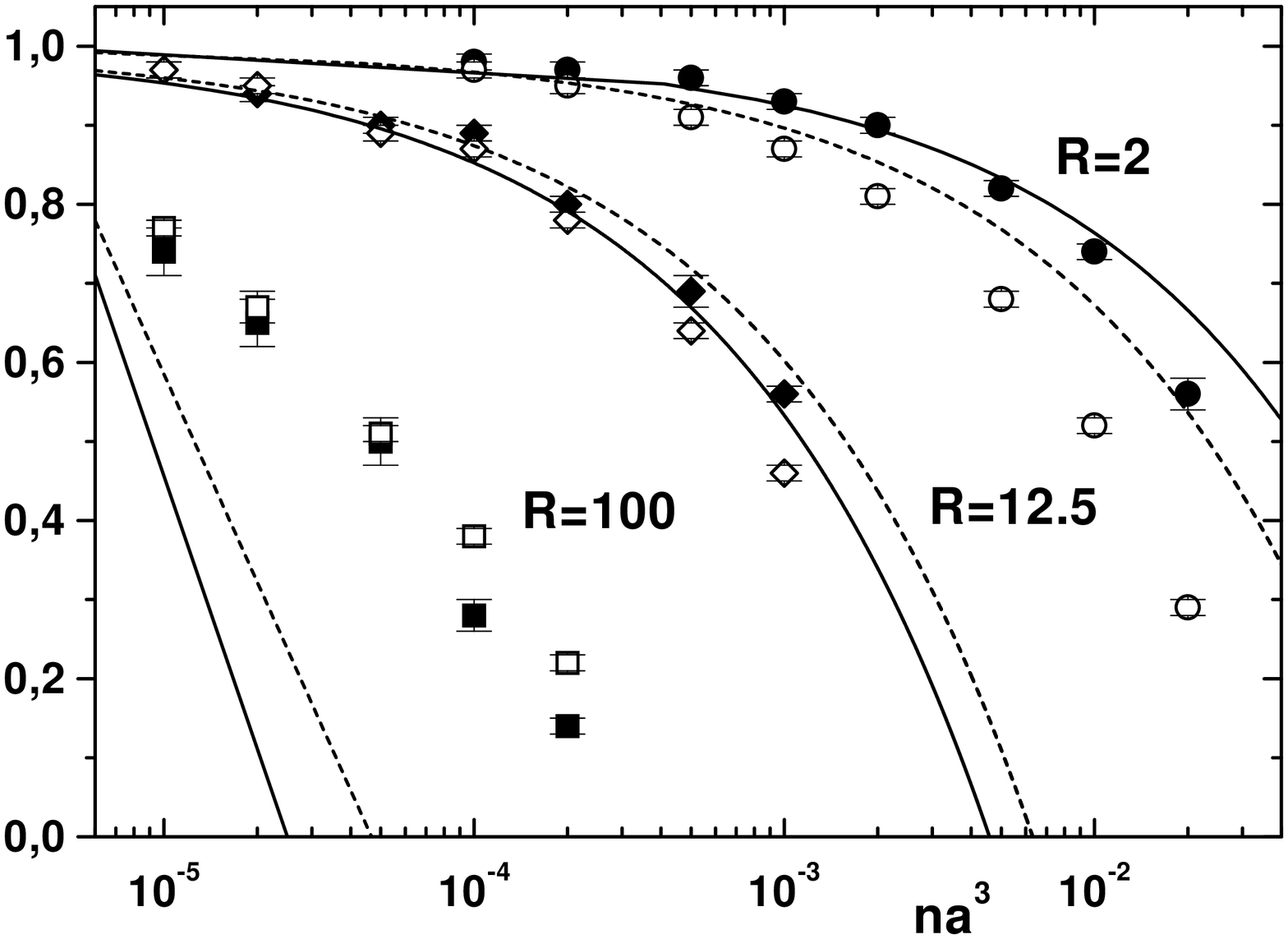}
\vspace{3 mm}
\caption{Superfluid fraction $\rho_s/\rho$ (solid symbols) and condensate fraction $N_0/N$ (open symbols).  
Disorder parameters are as in Fig.~\ref{Fig1}. Solid lines correspond to Eq. (\ref{SF}) and dashed lines to 
Eq. (\ref{CF}).}
\label{Fig2}
\end{center}
\end{figure}

\begin{figure}
\begin{center}
\includegraphics*[width=0.95\columnwidth,height=0.6\columnwidth]{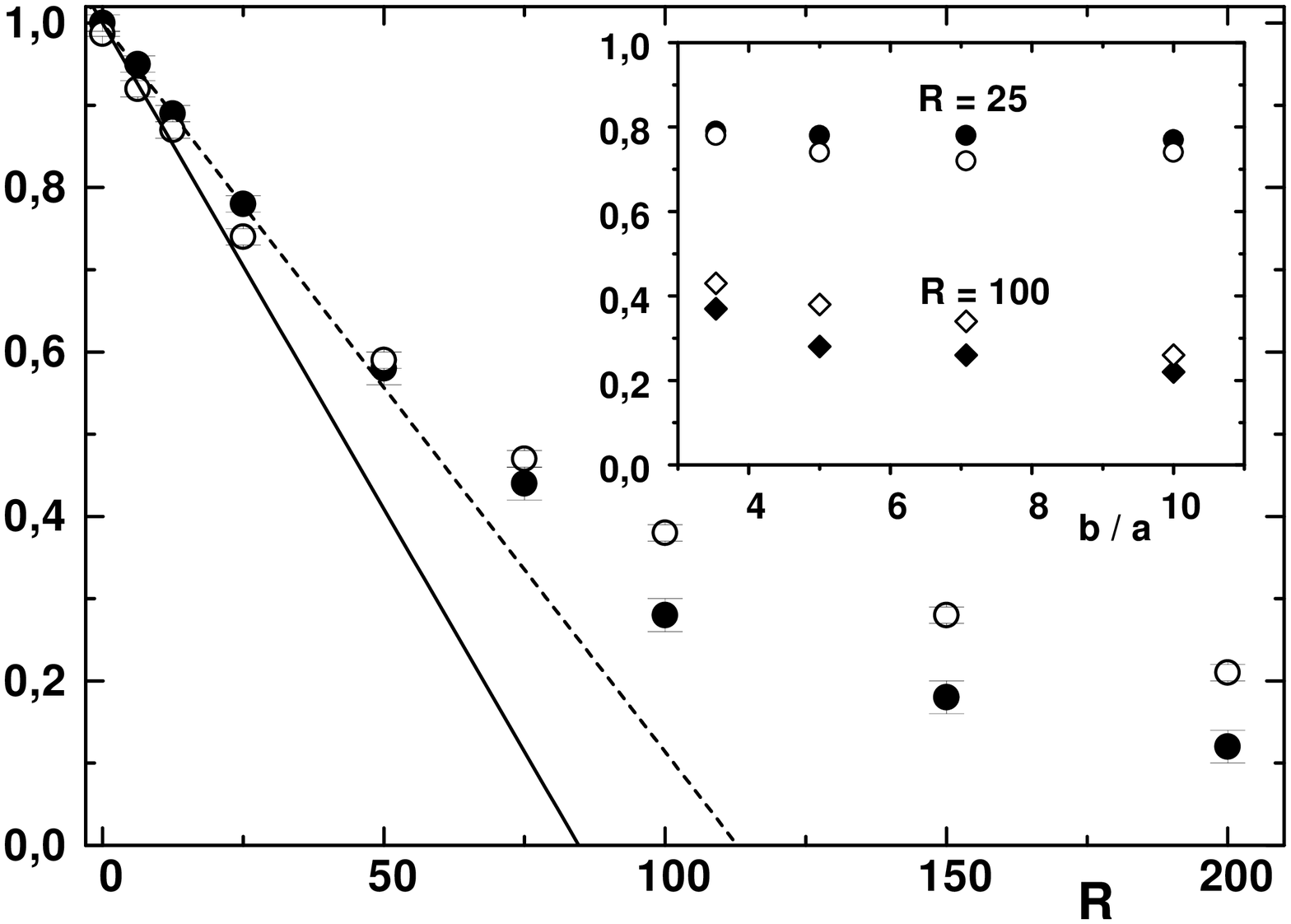}
\vspace{3 mm}
\caption{Superfluid fraction (solid symbols) and condensate fraction (open symbols) for $na^3=10^{-4}$. 
The strength of disorder $R$ has been varied by changing the concentration $\chi$ of impurities with a fixed 
ratio $b/a=5$. The solid line corresponds to Eq. (\ref{SF}) and the dashed line to Eq. (\ref{CF}). 
{\it Inset.} Scaling behavior as a function of the ratio $b/a$ for given values of the strength $R$. Error bars
have approximately the size of the symbols.}
\label{Fig3}
\end{center}
\end{figure}

\begin{figure}
\begin{center}
\includegraphics*[width=0.95\columnwidth,height=0.6\columnwidth]{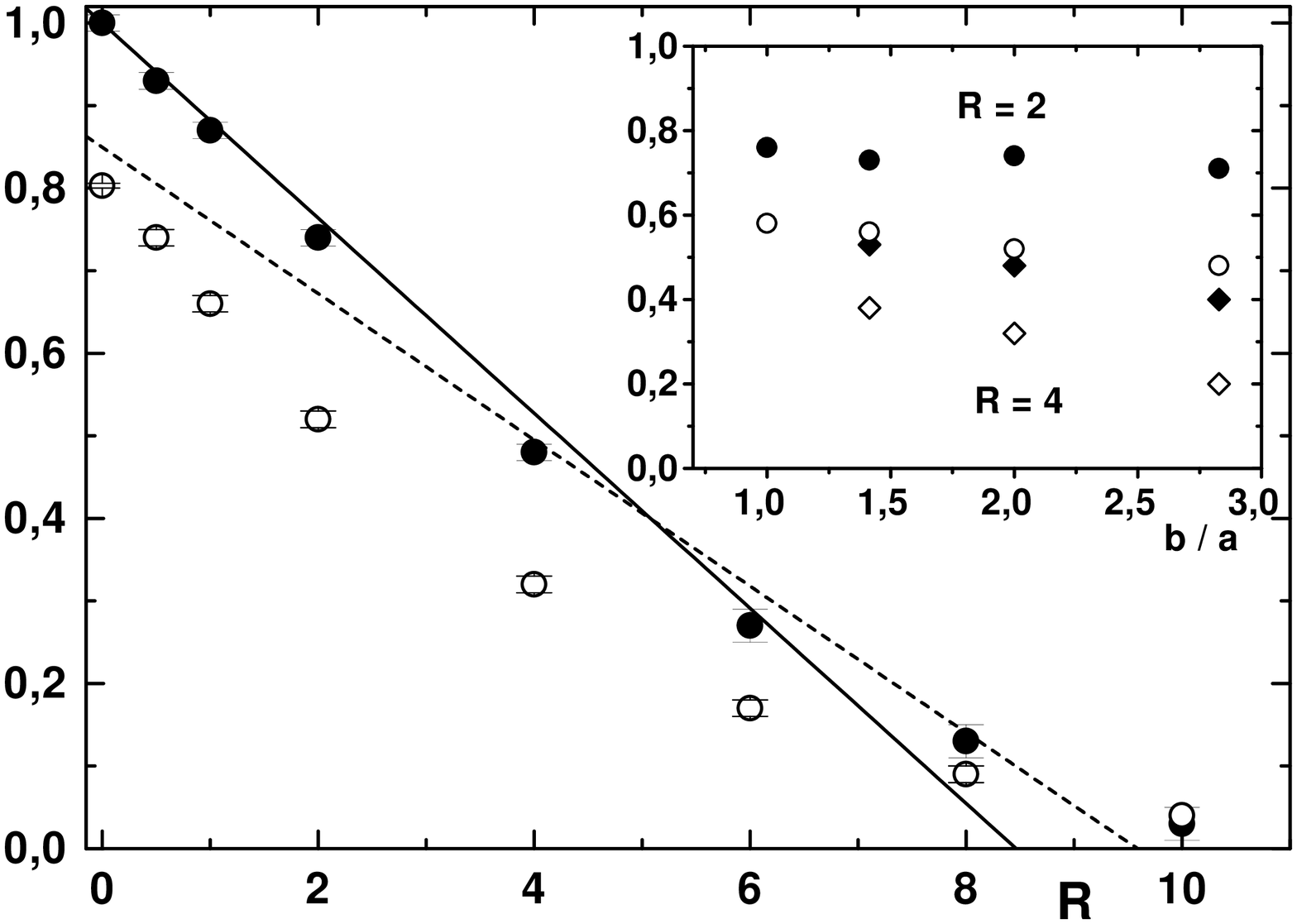}
\vspace{3 mm}
\caption{Superfluid fraction (solid symbols) and condensate fraction (open symbols) for $na^3=10^{-2}$. 
The strength of disorder $R$ has been varied by changing the concentration $\chi$ of impurities with a fixed 
ratio $b/a=2$. The solid line corresponds to Eq. (\ref{SF}) and the dashed line to Eq. (\ref{CF}). 
{\it Inset.} Same as in Fig.~\ref{Fig3}.}
\label{Fig4}
\end{center}
\end{figure}

{\it Results.} In Fig.~\ref{Fig1}, results for the energy beyond mean field as a function of
the gas parameter and for different strengths of disorder are presented. 
For $R=2$ we find good agreement with Eq. (\ref{gse})  
over a wide range of densities. By increasing $R$, deviations start to appear at lower 
densities. In particular, for the largest value $R=100$, we do not find agreement for densities larger than 
$na^3>10^{-5}$. 
In Fig.~\ref{Fig2}, we show results for $\rho_s/\rho$ and $N_0/N$. For $R=2$ the superfluid fraction 
follows the analytical 
prediction (\ref{SF}) up to large values of $na^3$. On the contrary, the condensate fraction is more
sensitive to the increase of density and deviates earlier from the Bogoliubov result (\ref{CF}). 
The value $R=12.5$ corresponds to a strength of disorder above the critical value ($R_c=5.1$), where the Bogoliubov 
model predicts $\rho_s/\rho<N_0/N$. We do not see this behavior.  In fact, although the agreement between
$\rho_s/\rho$ and Eq. (\ref{SF}) is good up to relatively large values of $na^3$, the 
depletion of the condensate becomes very soon larger than predicted by Eq. (\ref{CF}) and, as a consequence, 
we find either $\rho_s/\rho\simeq N_0/N$ at very low densities or $\rho_s/\rho>N_0/N$ for larger densities.
The results for $R=100$ correspond to a regime of strong disorder where Bogoliubov model can not be applied. 
In this regime, $\rho_s/\rho$ and $N_0/N$ first decrease together with increasing density and then, for 
$na^3\ge 10^{-4}$, a clear gap appears with the superfluid fraction significantly smaller than the condensate fraction.
To our knowledge this is the first direct realization of a system exhibiting this unusual feature.  
The cross-over from weak to strong disorder is better shown in Fig.~\ref{Fig3}. In the figure 
we present results for $\rho_s/\rho$ and $N_0/N$ as a function of $R$ at the density $na^3=10^{-4}$. 
By increasing the strength of 
disorder, superfluid and condensate fractions first decrease together, and for large values of $R$
the strong disorder regime of Fig.~\ref{Fig2} where $\rho_s/\rho<N_0/N$ is achieved. At large density the 
situation is different as shown in Fig.~\ref{Fig4}, where $na^3=10^{-2}$. 
Already in the absence of disorder interaction effects give rise to a sizable depletion of the condensate
(about 20\%) and by adding disorder no clear evidence of a regime where $\rho_s/\rho<N_0/N$ is observed.
An interesting result which emerges from Figs.~\ref{Fig2}-\ref{Fig4} is that the behavior of the superfluid fraction 
is well described by the Bogoliubov prediction (\ref{SF}) also for high densities, provided $R$ is small.
On the contrary, the condensate fraction is much more sensitive to the value of the gas parameter and agreement with 
(\ref{CF}) is found only in the regime where both $na^3<<1$ and $\sqrt{na^3}R<<1$. 

In the regime where $N_0/N$ and $\rho_s/\rho$ agree with the analytical predictions [results (\ref{CF}), (\ref{SF})], 
the scaling behavior on the parameter $R$ is evident. An important result of our analysis concerns the fact that 
the scaling behavior extends well beyond the region where results (\ref{CF}) and (\ref{SF}) apply. 
This is explicitly shown in the insets of Fig.~\ref{Fig3}
and \ref{Fig4}, where we vary both the ratio $b/a$ and the concentration $\chi$ with $R=\chi\;(b/a)^2$ fixed. 
At small density (Fig.~\ref{Fig3}) we find that, even in the case of strong disorder $R=100$, deviations from 
scaling are relatively small. At large density (Fig.~\ref{Fig4}) we still find good scaling for $R=2$, whereas 
for $R=4$ a dependence on the value of $b/a$ becomes evident.
Due to the constraint of 
non-overlapping impurities systems with larger strengths of disorder can not be studied. 
Nevertheless, we have investigated the occurrence of a quantum phase transition by analyzing the dependence 
of the results for 
$\rho_s/\rho$ and $N_0/N$ on the size of the system. Our DMC calculations show no 
significant finite-size effects and the results shown in Figs.~\ref{Fig3},\ref{Fig4} are thus appropriate to the 
thermodynamic 
limit. We conclude that within our model of non-overlapping impurities there is no quantum phase transition 
for a critical value of disorder.  

In conclusion, we have investigated BEC and superfluidity in a Bose gas with disorder as a function of density
and strength of disorder. We have shown that dilute systems with weak disorder can be correctly described using 
the Bogoliubov model. For strong disorder we find that the system exhibits the unusual feature of a superfluid 
fraction significantly smaller than the condensate fraction, in qualitative agreement with the prediction of 
Bogoliubov model.  

The authors would like to thank L.P. Pitaevskii and S. Stringari for many useful discussions. One of us (S.G.)
acknowledges the hospitality of the Aspen Center for Physics. This research has been partially supported by DGES
(Spain) Grant No. PB98-0922. We also acknowledge supercomputer facilities provided by CEPBA.

\end{document}